\documentclass[10pt,a4paper,conference]{IEEEtran}

\IEEEoverridecommandlockouts

\usepackage[nolist]{acronym}                    
\usepackage{amssymb}
\usepackage{paralist}
\usepackage{multirow}
\usepackage{microtype}
\usepackage{booktabs}                   

\ifCLASSINFOpdf
  \usepackage[pdftex]{graphicx}
\else
\fi

\usepackage[cmex10]{amsmath}
\allowdisplaybreaks

\usepackage{array}

\usepackage{mdwtab}

\usepackage[caption=false,font=footnotesize]{subfig}

\usepackage{fixltx2e}

\usepackage{url}

\begin{document}

\title{Specifying and Placing Chains of\\ Virtual Network Functions\vspace{-8mm}}

\author{\IEEEauthorblockN{Sevil Mehraghdam\thanks{This work is partially supported 
by the International Graduate School ``Dynamic Intelligent Systems''.}}
\IEEEauthorblockA{University of Paderborn\\
33098 Paderborn, Germany\\
Email: s.mehraghdam@uni-paderborn.de}
\and 
\IEEEauthorblockN{Matthias Keller}
\IEEEauthorblockA{University of Paderborn\\
33098 Paderborn, Germany\\
Email: mkeller@uni-paderborn.de}
\and
\IEEEauthorblockN{Holger Karl}
\IEEEauthorblockA{University of Paderborn\\
33098 Paderborn, Germany\\
Email: holger.karl@uni-paderborn.de}}

\maketitle

\begin{abstract}
Network appliances perform different functions on network flows and constitute an 
important part of an operator's network. Normally, a set of chained network functions
process network flows. Following the trend of virtualization of networks, 
virtualization of the network functions has also become a topic of interest.  
We define a model for formalizing the chaining of network functions using a 
context-free language. We process deployment requests and construct virtual
network function graphs that can be mapped to the network. We describe the mapping as a Mixed 
Integer Quadratically Constrained Program (MIQCP) for finding the placement of 
the network functions and chaining them together considering the limited network resources 
and requirements of the functions. We have performed a Pareto set analysis to 
investigate the possible trade-offs between different optimization objectives.
\end{abstract}

\IEEEpeerreviewmaketitle

\section{Introduction}
\label{sec:introduction}
\acresetall

An operator's network consists of a large number of intermediate \acp{nf}. 
\acp{nat}, load balancers, firewalls, and \acp{ids} are
examples of such functions.
Traditionally, these functions are implemented on physical middle-boxes, which are 
network appliances that perform functions other
than standard path discovery or routing decisions for forwarding packets (RFC~3234 \cite{Carpenter2002}).
Middle-boxes are based on special-purpose hardware platforms
that are expensive and difficult to maintain and upgrade. Following the trend 
of virtualization in large-scale networks, network function that were deployed as 
middle-boxes are also being replaced by \emph{Virtual} Network Functions (VNFs). \acused{vnf}

Typically, network flows go through several network functions. That means 
a set of \acp{nf} is specified and the flows traverse these \acp{nf} 
in a specific order so that the required functions are applied to the flows. This
notion is known as \emph{network function chaining} or \emph{network service chaining}
\cite{schneiderstandardizations, draft-ietf-sfc-problem-statement-05}.

\acp{nf} can modify the traversing network flows in different ways. 
For example, a \ac{dpi} can split the incoming flows over different branches 
according to the type of the inspected packets, each branch having a fraction of 
the data rate of the incoming flow. Firewalls can drop certain packets, 
resulting in flows with a lower data rate than incoming flows. A video optimizer 
can change the encoding of the video, which can result in a higher data rate. 
There can also be a dependency among a set of \acp{nf} that should be applied 
to the traffic in a network \cite{Sekar2011}, which requires special attention 
to the \emph{order} of traversing the functions in chaining 
scenarios. For instance, if the packets have to go through a WAN optimizer and an \ac{ids}, 
packet inspection by the \ac{ids} should typically be carried out before the WAN optimizer 
encrypts the contents. In case the functions in a chaining scenario do 
not have such a dependency, there can be multiple possibilities for chaining them 
together. Depending on how each function in the chain modifies the data rate of 
the flows, different chaining options can have different impact on the traffic 
in network links, on application performance, or on latency. We address two 
challenges in this area.  

The \emph{first challenge} is to formalize a request for chaining several \acp{nf} together, while considering
the possible dependencies among them. Upon receiving such a request, the network 
operator has the freedom to chain the functions in the best possible way to fit 
the requirements of the tenant applications. 
After the logical chaining of functions is specified, the functions need to be 
placed in the operator's network. In addition to the dependencies within the 
chains of functions, \acp{nf} can also be shared and reused among different 
applications in the network. The \emph{second challenge} is hence to find the best placement
for the functions, considering the requirements of individual requests as well as
the overall requirements of all network applications and the combination of requests.
We will elaborate on these considerations in the following sections. For example, 
Figure~\ref{fig:lb} illustrates two different ways of chaining a set of functions
together, one resulting in a lower average data rate requirement but higher processing 
requirements than the other. 

Traditionally, modifying the way functions are chained together and changing the
placement of functions in the network require complex modifications to the network, 
such as modifying the network topology or changing how physical boxes are connected
to each other. \ac{nfv} offers more flexibility in network function chaining by
simplifying chaining and placement of \acp{nf} and by 
making these changes more practical and achievable in real scenarios.  
Considering this, we focus on formalizing the chaining requests and placing the 
chained functions in the best locations according to optimization goals in an 
operator's network that supports \ac{nfv}. Our solutions are not bound to
a specific implementation option and can be applied to virtualized and non-virtualized
\ac{nf} chaining scenarios.

First, we give an overview of the related work in Section~\ref{sec:relatedwork}.
In Section~\ref{sec:specification}, we define our model for specifying the requests for 
chaining \acp{vnf} (given by network administrators or tenants) in a
way that different chaining options can be analyzed 
and the one that fits the optimization goals and network requirements can be
chosen. 
Based on the chaining possibilities given in the requests, the operator
can decide the placement of the functions. The preprocessing step required
before the functions can be placed in the network and a heuristic
for reducing the runtime of the final decision process are described in 
Section~\ref{sec:transformation}. We define a \ac{miqcp} in Section~\ref{sec:placement} 
for placing the chained \acp{vnf} considering the following three objectives:
\begin{inparaenum}[1)]
 \item Maximizing remaining data rate;
 \item minimizing number of used network nodes;
 \item minimizing total latency over all paths.
\end{inparaenum}
Finally, we describe the results of evaluating our placement and a multi-objective
Pareto discussion in Section~\ref{sec:evaluation} and conclude the paper in 
Section~\ref{sec:conclusion}.

\section{Related Work}
\label{sec:relatedwork}

A number of standardization activities in the \ac{nfv} area are carried out by 
\acs{etsi} and \acs{ietf}, resulting in a white paper~\cite{Introduction2012} and 
several drafts like problem statements in \ac{nfv}~\cite{draft-xjz-nfv-model-problem-statement-00, draft-ietf-sfc-problem-statement-05}, 
use cases~\cite{draft-liu-sfc-use-cases-05} and frameworks for network function chaining~\cite{draft-boucadair-sfc-framework-02}.
The model for chaining \ac{vnf} presented in this paper is compatible with the models and architectures proposed in these drafts.

Our placement solution for chained \acp{nf} can be considered as an extension to 
the following two NP-hard problems: Location-Routing Problems (LRP) and Virtual
Network Embedding (VNE). 
Location-routing problems \cite{Nagy2007,Prodhon2014} aim at placement of components 
while reducing the costs in nodes, edges or paths. 
In these problems, each path has one start point and one end point. We need to
create several paths between different pairs of \acp{nf} and connect these paths 
to represent the chaining. In this case, the routing problem turns into a multi-commodity 
flow problem, with inter-commodity dependencies.

Virtual network embedding (VNE) problems~\cite{Fischer2013} are similar to our problem 
in the chaining aspect. Chained \acp{nf} can be seen as graphs to be 
embedded into a substrate network. These problems treat the nodes of the virtual graphs 
(\acp{nf} in our case) independently. In our problem, however, flows from different 
tenants can share and reuse \acp{nf} for better \ac{nf} utilization.
Similar to the approach of Fuerst et al.~\cite{Fuerst2013}, we place several 
\acp{nf} on a single node to reduce inter-location traffic.

Joseph and Stoica~\cite{Joseph2008} present a model for describing the behavior 
and functionality of a limited number of \acp{nf} but it is not a complete model 
for \acp{vnf} since it does not contain any information about requirements like 
computational resources and their influence on the network traffic. 
Gember~et~al.~\cite{gember2013stratos} introduce a network-aware orchestration 
layer for \acp{nf}. Our model for chaining \acp{vnf} is similar in some
aspects, but their model is defined in a data center network, while we focus on an operator's network with multiple data center sites. 
Moreover, they do not capture any resource requirements of the functions 
apart from processed traffic. 

\ac{nfv} is still a concept under investigation and standardization of the problem 
definition and use cases are not yet finalized. Therefore, we found only a very 
limited amount of ongoing work related to our research focus. We define our models in 
a flexible way that can be extended and reinterpreted easily when more
technical details about the chaining scenarios and implementation requirements are
available.

\section{Network Function Chaining Specification}
\label{sec:specification}

We model the substrate network, where \ac{vnf} chains are defined and placed, 
as a connected directed graph, $ G {=} (V,E) $. 
Some of network nodes are \emph{switch nodes}, with typical routing and 
switching capabilities, along with (small) computational capacity that can be 
used for running \acp{vnf}, e.g., inside an FPGA in the switch. The remaining nodes 
are distributed sites with (much larger) computational capacity. We consider 
each of these sites as a large computational unit, called a 
\emph{data center node}, without looking into their internal topology. 
We define two types of computational capacities for the nodes: 
$ c_d(v) $ and $ c_s(v) $ $ (\forall v {\in} V) $. For today's switch nodes, $ c_d $
is zero and for current data center nodes, $ c_s $ is zero. We define 
both types of capacities for all nodes to keep our model open to future extensions. 
For example, in future, switches might be equipped
with general-purpose processing capabilities, leading to $ c_d > 0 $ for switch nodes.
The network links are directed edges in the graph, with data rate 
$ d(v,v') $ and latency $ l(v,v') $ for every edge $ (v,v') {\in} E $.


The following information about offered network functions is available and 
maintained by the network operator:
\begin{itemize}
 \item Set $ F $ of available network functions.
 \item Computational resource requirements $ p(f) \, (\forall f \in F) $ of an 
 instance of the \ac{vnf} $ f $ per each request, whether it is placed on a switch 
 node, $ p_s(f) $, or on a data center node, $ p_d(f) $.
 Some functions can be placed either on a switch or on a data center node, e.g., 
 a load balancer ($ p_d {>} 0 $ and $ p_s {>} 0 $), and some can be placed only 
 on a data center node, e.g., a \ac{vm} implementing a video optimizer 
 ($ p_d {>} 0 $ and $ p_s {=} 0 $).
 \item Maximum number of instances of the \ac{vnf} that can be deployed, 
 $ n_{\text{inst}}(f) $, e.g., determined by the number 
 of licenses that the operator owns for the \ac{vnf}.
 \item Number of chaining requests an instance of a \ac{vnf} can handle, 
 $ n_{\text{req}}(f) $. For example, 
 an anti-virus function can be configured once and used for every chain that needs this function,
 (i.e., number of requests it can handle is only limited by hardware specifications),
 but a firewall might need specific configurations for each chaining request and
 one instance of this function cannot be shared between two chains (i.e., $ n_{\text{req}} {=} 1 $).
\end{itemize}

A network operator receives \emph{deployment requests} for different partially ordered sets 
of \acp{vnf}. In these requests, a network administrator or the tenant specifies which 
of the offered functions should be applied in which order to given flows
between fixed start and end points. A deployment request contains the following information:
\begin{itemize}
 \item Set $ U $ of individual requests for instances of available network functions.
 \item \emph{Chaining request}, denoted as $ c $, for specifying the desired order of functions.
 \item For each branch leaving a requested function, the percentage of incoming
 data rate it produces given as an ordered set $ r(u) \, (\forall u {\in} U) $ for each function.
 For example, for a \ac{dpi} that is expected to send 20\% of the incoming packets 
 towards a video optimizer and 80\% towards a firewall this set is given as $ \{ 0.2,0.8 \} $.
 \item Set $ A $ of fixed start and end points for the flows, e.g., an application 
 \ac{vm} deployed in a data center node or 
 a router that connects the operator's network to external networks. 
 \item Set $ A_{\text{pairs}} {\subseteq} A {\times} A $ of \emph{pairs} of start 
 and end points belonging to different flows.
 \item Location of start and end points of flows in the network, $ \text{loc}(a) {\in} V \, (\forall a {\in} A) $.
 \item Initial data rate entering the chained functions, $ d_{\text{in}} $.
 \item Maximum tolerable latency between the start and end points of flows, 
 $ l_{\text{req}}(a,a') \, (\forall (a,a') {\in} A_{\text{pairs}}) $.
\end{itemize}

We define a context-free language for formalizing the chaining requests. Using this 
language, complex requests can be composed that contain the order of functions to 
be applied to network flows. Every chaining request is 
formalized using different types of \emph{modules}. The elements of this language 
are the following modules:
\begin{itemize}
 \item An individual function or a start/end point for the chain
 \item A set of functions that should be applied to network flows in
 an optional order (\emph{optional order} module)
 \item A function that splits the incoming flows into multiple branches
 that consist of different modules (\emph{split} module)
 \item A function that splits the incoming flows into multiple branches 
 that all consist of the same module (\emph{parallel} module)
\end{itemize}

Several (possibly nested) modules can be placed sequentially in the chaining request
to reflect a simple and fixed \emph{order} among desired functions.
 
$ \mathcal{G} {=} (\mathcal{V},\mathcal{T},\mathcal{P},\mathcal{S}) $ is the grammar
for this language. 
$ \mathcal{V} $ is the set of non-terminals consisting of the following variables:
$ \langle\text{modules}\rangle $, $ \langle\text{mod}\rangle $,
$ \langle\text{order}\rangle $, $ \langle\text{optorder}\rangle $, $ \langle\text{split}\rangle $,
$\langle\text{parallel}\rangle $, $ \langle\text{term}\rangle $, $ \langle\text{moremod}\rangle $,
$ \langle\text{moreterm}\rangle $, $ \langle\text{num}\rangle $.

$ \mathcal{T} {=}  U {\cup} A {\cup} D {\cup} \{1,2,\dotsc,n\} {\cup} \{\epsilon\} $ is the
set of symbols of this language where $ \epsilon $ is the empty string. 
A subset of natural numbers from $ 1 $ to $ n $ is 
required for displaying the number of branches that leave a \ac{vnf}. $ n $ can be
defined as a number larger than the maximum number of outgoing branches in all requests. 
This upper bound is necessary because the set of symbols has 
to be a finite set. $ D $ is the set of delimiters consisting of the following symbols:
. , ; \{ \} ( ) [  ]. $ \mathcal{S} {=} \langle\text{start}\rangle $ is the start 
symbol of the grammar and $ \mathcal{P} $ is the set of production rules:
{\footnotesize
  \begin{align}
  \label{rule:start} \langle\text{start}\rangle & ::= \langle\text{modules}\rangle \\ 
  \label{rule:modules}\langle\text{modules}\rangle & ::= \langle\text{order}\rangle~\langle\text{modules}\rangle ~|~\langle\text{mod}\rangle \\ 
  \label{rule:order}\langle\text{order}\rangle & ::= \langle\text{mod}\rangle~\cdot \\ 
  \label{rule:mod}\langle\text{mod}\rangle & ::= \langle\text{optorder}\rangle~|~\langle\text{split}\rangle~|~\langle\text{parallel}\rangle~|~\langle\text{term}\rangle \\  
  \label{rule:optorder}\langle\text{optorder}\rangle & ::= ~(~\langle\text{term}\rangle~\langle\text{moreterm}\rangle~) \\ 
  \label{rule:split}\langle\text{split}\rangle & ::= \langle\text{term}\rangle~[~\langle\text{modules}\rangle~\langle\text{moremod}\rangle~] \\ 
  \label{rule:parallel}\langle\text{parallel}\rangle & ::= \langle\text{term}\rangle~\{~\langle\text{term}\rangle~\langle\text{moreterm}\rangle~;~\langle\text{modules}\rangle~;~\langle\text{num}\rangle~\} \\ 
  \label{rule:moreinst}\langle\text{moreterm}\rangle & ::= ~,~\langle\text{term}\rangle~\langle\text{moreterm}\rangle~|~\epsilon \\  
  \label{rule:moremod}\langle\text{moremod}\rangle & ::= ~,~\langle\text{modules}\rangle~\langle\text{moremod}\rangle~|~\epsilon \\ 
  \label{rule:term}\langle\text{term}\rangle & ::= ~u_1~|~u_2~|~\dotsc~|~u_{|U|}~|~a_1~|~a_2~|~\dotsc~|~a_{|A|} \\ 
  \label{rule:num}\langle\text{num}\rangle & ::= ~1~|~2~|~3~|~\dotsc~|~n
  \end{align}
  }%

Rule~\ref{rule:mod} expresses the 4 different types of modules. We refer to the 
requests for using an instance of a \ac{vnf} and the start/end points of chains as 
\emph{terms} in this grammar. Rule~\ref{rule:order} is used for defining a fixed 
and simple order among modules. Optional order modules are produced by 
Rule~\ref{rule:optorder}. Such a module consists of a set of functions that
should be applied to the flows and the order of traversing these functions can be
chosen by the operator. A request for a split module can be expressed by 
Rule~\ref{rule:split}, where the splitting function is a \emph{term} and the modules
on different branches can be any of the defined types of modules. Finally, a request 
for a parallel module can be produced using Rule~\ref{rule:parallel}. Parallel 
modules have 4 parts:
\begin{inparaenum}
 \item The function that splits the flows into different branches; 
 \item a set of functions, including the splitting function, that can be placed 
 in an arbitrary order before the flows reach the modules on different branches 
 (optional);
 \item the module that should be replicated for the given number of times on
 multiple branches;
 \item number of outgoing branches from the splitting function (number of times 
 the module mentioned in part 3 should be replicated).
\end{inparaenum}

A \ac{vnf} chaining request formalized this way is a representation
for a connected directed graph that we refer to as a \emph{\ac{vnf} graph}. The required 
functions and also the start and end points of the flows are nodes of the \ac{vnf} 
graph ($ U {\cup} A $). The start/end points are mapped to fixed locations in the substrate
network and the location for the \acp{vnf} has be to determined. Each one of the 
directed links in the set of edges in this graph ($ U_{\text{pairs}}$) represents 
the order of traversing the functions. Every link in the \ac{vnf} graph has to be 
mapped to a path (consisting of at least of edge) in the 
substrate network graph. We define a two-step process for deploying the chained 
functions based on the deployment requests: processing the requests and building 
\ac{vnf} graphs, and finding the optimal placement for the \ac{vnf} graphs based 
on optimization goals in the operator's network. We describe this process in the following
sections.

\section{Processing Deployment Requests}
\label{sec:transformation}

The network operator receives deployment requests for placing chained \acp{vnf}
in the network, where chaining requests are formalized using the language described in 
Section~\ref{sec:specification}. The operator needs to find the best placement for several
chained functions based on multiple deployment requests. The first step is to build
\ac{vnf} graphs for every deployment request.

For each deployment request, the chaining request ($ c $) is first parsed using 
the grammar of the language (Section~\ref{sec:specification}). The parser 
matches and stores different modules of the request. Modules consisting of a single
function or start/end point of flows are stored as a node of the \ac{vnf} graph. 
For the modules where a number of functions can be ordered arbitrarily (\emph{optional order} and 
\emph{split} modules) every possible permutation of the set of functions is computed
and stored separately as a candidate for being a part of the final \ac{vnf} graph.
Moreover, for every match of a \emph{parallel} module, the module on 
the branches is replicated for the requested number of times and stored as a part
of the graph. Using the specified orders and depending on the modules that build 
the chaining request, at the end of the parsing process different modules are stored
as parts of the \ac{vnf} graph with explicit orders among all functions. Parts
of the graphs are then connected using directed links that represent the sequence 
of modules in the request. Considering the different permutations for different 
modules, at least one \ac{vnf} graph is built out of each chaining request. 
Using the rest of the information in the deployment request and the information 
available about the network functions, computational resource requirements are 
assigned to the nodes of the \ac{vnf} graphs. The links of the graphs are also
annotated by data rate and latency requirements. 

\begin{figure}[!t]
\subfloat[]{\includegraphics[width=0.45\linewidth]{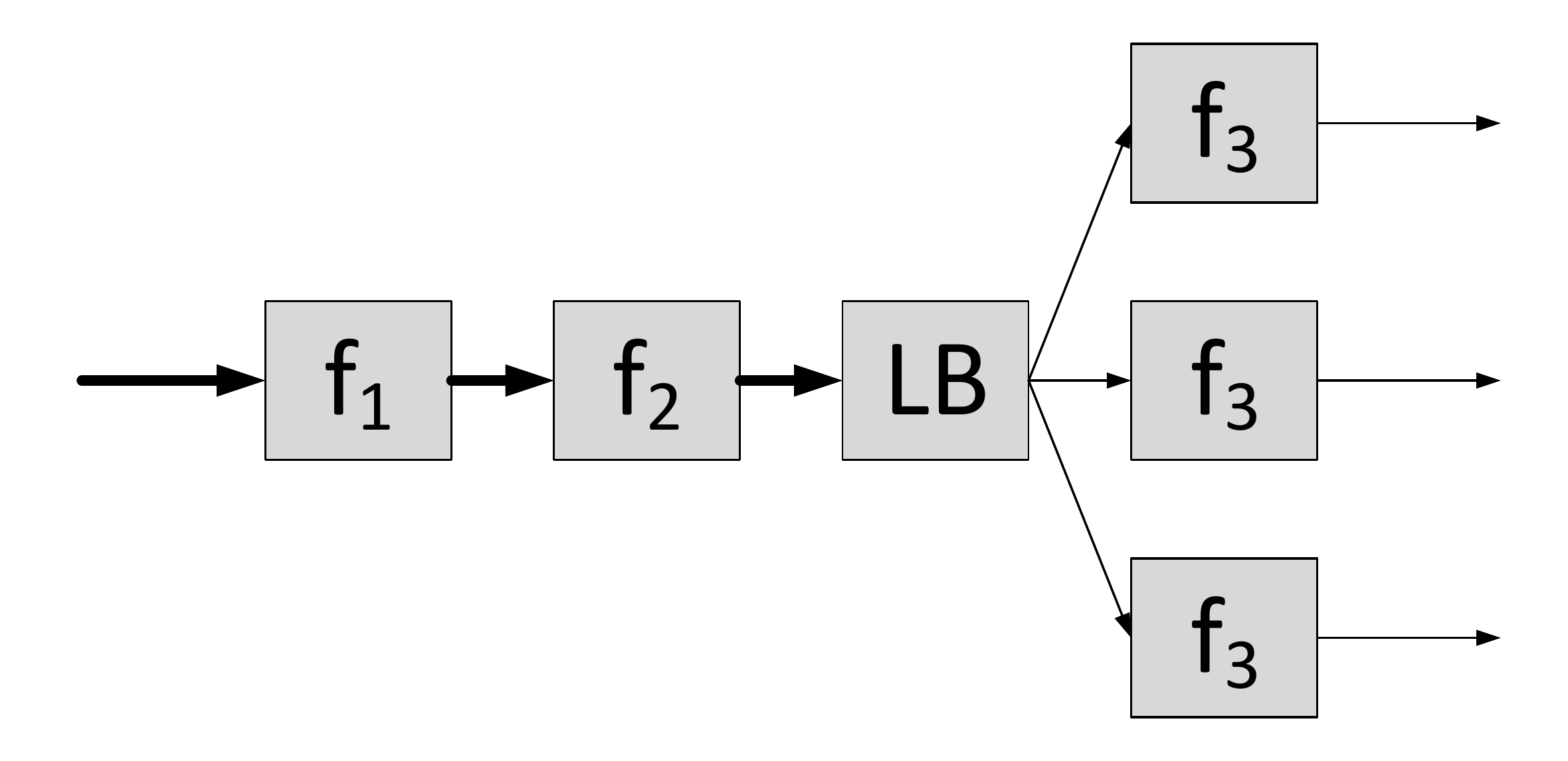}%
\label{fig:lb1}}
\hfil
\subfloat[]{\includegraphics[width=0.45\linewidth]{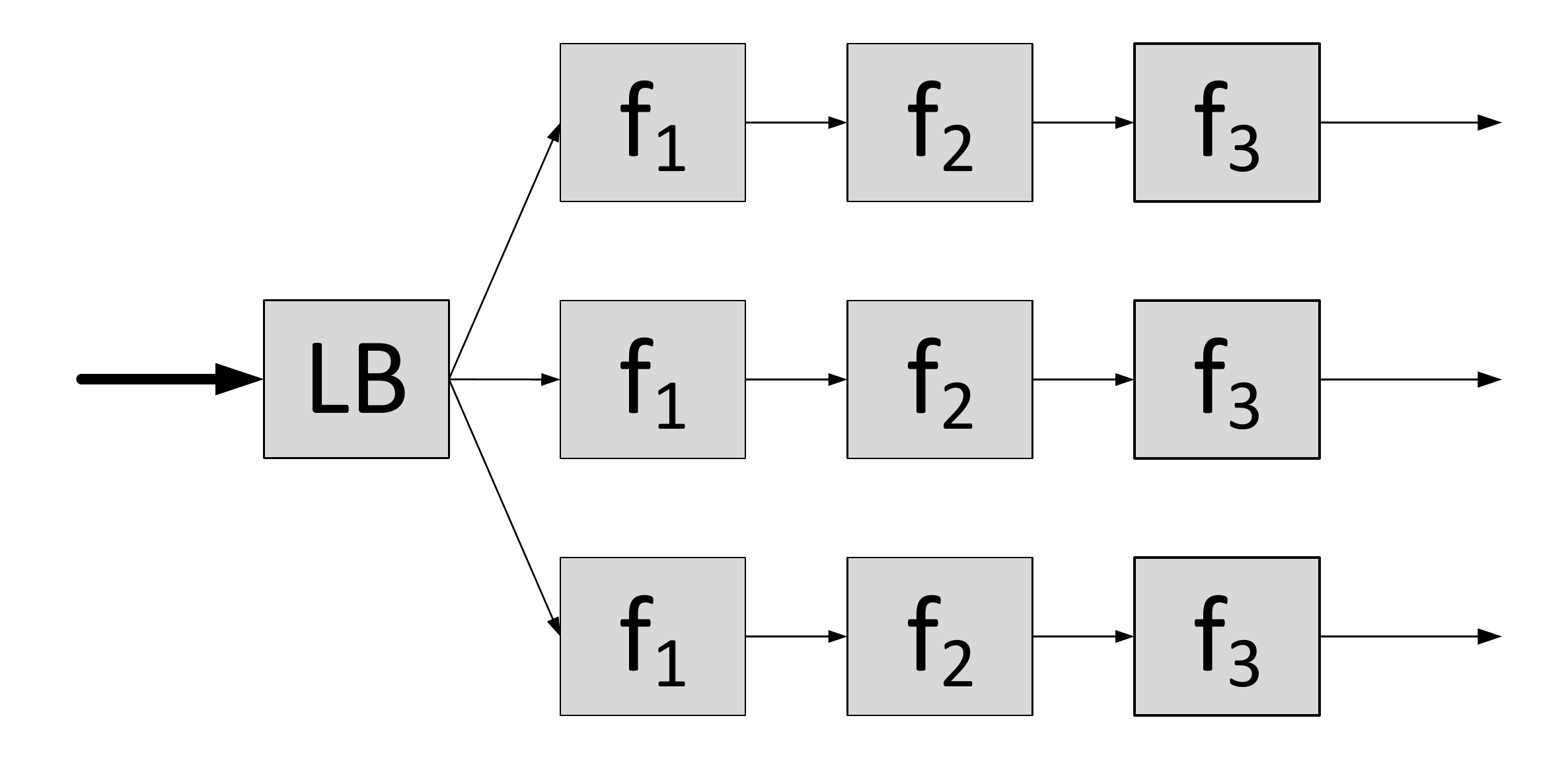}
\label{fig:lb2}}
\caption{Two of the chaining options for a set of functions}
\label{fig:lb}
\vspace{-3mm}
\end{figure}

Each of the graphs that can be created from a request can have different characteristics 
in terms of average data rate required for its links and number of \ac{vnf} instances. 
For example, the \ac{lb} in Figure~\ref{fig:lb} splits the incoming flows into 
three different branches to balance 
traffic over three instances of $ f_3 $. The ratio of outgoing to incoming data 
rate in all \acp{vnf} except the \ac{lb} is 1. Placing this load balancer earlier in
the directed graph, as in Figure~\ref{fig:lb2}, reduces the data rate of the 
links on each branch after it. But it also means that up to three instances of
all subsequent \acp{vnf} will be required on the paths towards $ f_3 $. Each of
these instances has a lower processing requirement than the instances
in Figure~\ref{fig:lb1} that should handle higher data rates.

For every module that requests $ n {\in} \mathbb{N}_{\geq0} $ \acp{vnf} with 
optional order among them (i.e., \emph{optional order} or \emph{parallel} module), 
$ n! $ permutations are computed and stored. For a 
chain that contains multiple optional order or parallel modules, the number of 
different ways for combining these modules is the product of the number of 
permutations for different modules. For example, if a chaining request contains 
one optional order module with 3 different \acp{vnf} and one parallel module in which 
4 \acp{vnf} can be placed with an arbitrary order, a total of $ 3! {\cdot} 4! {=} 144 $ 
combinations of these modules is possible. That means, for finding the best combination,
the placement step in the deployment process has to be done 144 times so that the 
results can be compared and the option that fits the network requirements can be
chosen. Our deployment process creates the \ac{vnf} graphs for each deployment 
request separately, and then all possible combinations of \ac{vnf} graphs from 
different requests would have to be computed and sent to the placement step. As the number 
of combinations increases very quickly in the number of deployment requests and  
the number of functions in each request, trying every possible combination
becomes impractical. 

We propose a heuristic for choosing one of the possible combinations, respecting
the optimization goals in an operator's network. In this method, instead of computing
all possible permutations of the sets of functions with arbitrary orders, we 
sort the functions in ascending order according to their ratio of outgoing to
incoming data rate. The function that reduces the data rate of the flows the most 
is placed before all other functions in the module. Therefore, each deployment 
request results in one single \ac{vnf} graph and the final input to the placement
step is a disconnected graph consisting of the \ac{vnf} graphs from different requests.
We define the placement optimization problem in Section~\ref{sec:placement} for 
finding the best placement in the substrate network for the combined \ac{vnf} graphs.
We optimize the placement for a new set of deployment requests by mapping all new 
requests together into a network that may or may not contain a previous deployment of chained 
functions. This can be extended to adapt the existing deployments to the new 
state of the network.

For each deployment request, our heuristic chooses a \ac{vnf} graph that has the 
minimum overall data rate requirement among all possible \ac{vnf} graphs for that
request.
This method might discard some graphs that are optimal 
in terms of total number of required \ac{vnf} instances or the total latency. However,
the gain in execution time can compensate for this deviation from optimality.
We have performed a Pareto analysis of the placement optimization problem to show
the trade-offs between different optimization objectives, which we present in 
Section~\ref{sec:evaluation}.

\section{Placement of Chained Network Functions}
\label{sec:placement}

There can be several
metrics that the operator might need to optimize. We formulate the placement
optimization problem as an \ac{miqcp} with respect to data rate, number 
of used network nodes, and latency. Input to the placement step is the capacity
of network nodes and links, requirements of different network functions, and the
combined \ac{vnf} graph from the request processing step. Table~\ref{table:placementparams}
shows an overview of the input parameters to the placement optimization problem.

\begin{table}[!t]
\caption{Required input for placement}
\label{table:placementparams}
\centering
\begin{tabular}{lll}
\hline
Domain                                                   & Parameter               & Description                         \\ \hline\hline
\multirow{2}{*}{$ \forall v {\in} V $}                     & $ c_d(v) $              & Data center computational resources in $ v $          \\  
\multicolumn{1}{l}{}                                   & $ c_s(v)$               & Switch computational resources in $ v $              \\ \hline
\multirow{2}{*}{$ \forall (v,v') {\in} E $}                & $ d(v,v') $             & Data rate capacity on $ (v,v') $                          \\  
                                                         & $ l(v,v') $             & Latency of $ (v,v') $                                       \\ \hline
\multirow{4}{*}{$ \forall f {\in} F $}                     & $ n_\text{inst}(f) $    & Number of allowed instances for $ f $                 \\  
                                                         & $ n_\text{req}(f) $     & Number of requests $ f $ can handle       \\  
                                                         & $ p_d(f) $              & Data center resource demand of $ f $    \\  
                                                         & $ p_s(f) $              & Switch resource demand of $ f $        \\ \hline
$ \forall u {\in} U $                                      & $ t(u) $                & Requested function               \\ \hline
$ \forall (u,u') {\in} U_{\text{pairs}} $                  & $ d_\text{req}(u, u') $ & Data rate demand of $ (u,u') $                            \\ \hline
$ \forall a {\in} A $                                      & $ \text{loc}(a) $       & Network node where $ a $ is placed \\ \hline
\multirow{2}{*}{$ \forall (a,a') {\in} A_{\text{pairs}} $} & $ \text{paths}(a, a') $ & All possible paths between $ a $ and $ a' $        \\  
                                                         & $ l_\text{req}(a,a') $  & Maximum latency between $ a $ and $ a' $  \\ \hline
\end{tabular}
\end{table}

Decision variables are described in Table~\ref{table:variables}. 
``remdr'' and ``lat'' are continuous variables and all other ones are binary 
indicator variables. We show the constraints of the optimization problem in 
Section~\ref{subsec:constraints} and the objective functions in Section~\ref{subsec:objective}.

\begin{table}[!t]
\caption{Decision variables}
\label{table:variables}
\centering
\begin{tabular}{lll}
\hline
Domain                                                                                                                     & Variable 				 	& Description                                                                                                                                                 \\ \hline\hline
\multirow{3}{*}{\begin{tabular}[c]{@{}c@{}}$ \forall u {\in} U,$ \\ $ \forall v {\in} V $\end{tabular}}                                    & $ m_{u,v} $         	& $ u $ mapped to $ v $                                                                                                              \\  
\multicolumn{1}{l}{}                                                                                                     & $ \text{ms}_{u,v} $     	& $ u $ mapped to a switch function on $ v $                                                                                         \\  
                                                                                                                           & $ \text{md}_{u,v} $     	& $ u $ mapped to a data center function on $ v $                                                                                    \\ \hline
\begin{tabular}[l]{@{}c@{}}$ \forall f {\in} F, $ \\ $ \forall v {\in} V $\end{tabular}                                        & $ i_{f,v} $        	& An instance of $ f $ mapped to $ v $                                                                                                  \\ \hline
\begin{tabular}[l]{@{}l@{}}$ \forall (v,v') {\in} E, $ \\ $ \forall x,y {\in} V, $ \\ $ \forall (u,u') {\in} U_\text{pairs} $ \end{tabular} & $ e_{v,v',x,y,u,u'} $   		&  \begin{tabular}[l]{@{}l@{}}$ (v,v') $ belongs to path between $ x $ and $ y $, \\ where $ u $ and $ u' $ are mapped to\end{tabular} \\ \hline
$ \forall v {\in} V $                                                                                                        & $ \text{used}_{v} $     	& At least one request mapped to $ v $                                                                                         \\ \hline
$ \forall (v,v') {\in} E $                                                                                                   & $ \text{remdr}_{v,v'} $ 	& Remaining data rate on $ (v,v') $                                                                                                                    \\ \hline
$ \forall (u,u') {\in} U_\text{pairs} $                                                                                      & $ \text{lat}_{u,u'} $   	& Latency of the path between $ u $ and $ u' $                                                                                          \\ \hline
\end{tabular}
\end{table}

\subsection{Constraints}
\label{subsec:constraints}

In this section, we describe the constraints of the placement optimization problem 
in 3 parts:
\begin{inparaenum}
 \item Placing functions in network nodes and mapping requests for using instances
 of network functions to these nodes;
 \item creating paths between functions;
 \item collecting metric values.
\end{inparaenum}
Placement and path creation constraints have a clear separation that facilitates 
extending the model in either part without causing problems in the other part.
Besides, necessary ties between these parts are also carefully defined, for example, using
the decision variable $ e $, to build a
consistent and uniform model for placing functions and chaining them together optimally.

\subsubsection{Network Function Placement Constraints}

{\footnotesize
\begin{align}
   & \forall u {\in} U : \sum_{v {\in} V} m_{u,v} {=} 1 \label{constr:m_{u,v}} \\
   & \forall a {\in} A : m_{a,\text{loc}(a)} {=} 1 \label{constr:m_a} \\
   & \forall f {\in} F, \forall v {\in} V : \notag\\
   &\quad \sum_{u {\in} U  t(u) {=} f} m_{u,v} {\leq} \mathcal{M} {\cdot} i_{f,v} \label{constr:m_uv<=}  \\
   &\quad i_{f,v} {\leq} \sum_{u {\in} U, t(u) {=} f} m_{u,v} \label{constr:i_fv<=} \\
   & \forall u {\in} U, \forall v {\in} V : \text{ms}_{u,v} {+} \text{md}_{u,v} {=} 1 \label{constr:msmd} \\ 
   & \forall u {\in} U, \forall v {\in} V, p_s(t(u)) {=} 0, p_d(t(u)) {\neq} 0 : \notag\\
   &\quad \text{ms}_{u,v} {=} 0 \label{constr:ms} \\
   &\quad \text{md}_{u,v} {=} 1 \label{constr:md} \\
   & \forall v {\in} V : \notag\\
   &\quad \sum_{u {\in} U} m_{u,v} {\cdot} \text{md}_{u,v} {\cdot} p_d(t(u)) {\leq} c_d(v) \label{constr:m.md.p} \\
   &\quad \sum_{u {\in} U} m_{u,v} {\cdot} \text{ms}_{u,v} {\cdot} p_s(t(u)) {\leq} c_s(v) \label{constr:m.ms.p} \\
   & \forall f {\in} F : \sum_{v {\in} V} i_{f,v} {\leq} n_{\text{inst}}(f) \label{constr:i<=inst} \\
   & \forall v {\in} V, \forall f {\in} F : \sum_{u {\in} U, t(u) {=} f} m_{u,v} {\leq} n_{\text{req}}(f) \label{constr:m<=req} 
\end{align}
}%
Every request for using a \ac{vnf} should be mapped to exactly one node (Constr.\,\ref{constr:m_{u,v}}). Start/end points
of the flows are fixed in the network, so $ \forall a {\in} A $, $ m_{a,\text{loc}(a)} $
is not a decision variable. However, values of $ m $ are also used in path creation 
constraints and there are paths to be created between the start/end points and 
the rest of the chained functions. Therefore, for these points $ m $ is defined 
similar to the functions (Constr.\,\ref{constr:m_a}). Constraints \ref{constr:m_uv<=}
and \ref{constr:i_fv<=} are complementary to each other and avoid starting an
instance of a function on network nodes without mapping any requests to them. They
also make sure that a request is mapped to a node only if an instance of the 
required function is placed on that node. $ \mathcal{M} {\in} \mathbb{N} $ is a 
number larger than the sum on the left side of the inequality in Constr.\,\ref{constr:m_uv<=}
(a so-called ``big~M'' constraint). Constr.\,\ref{constr:msmd} ensures that a request is mapped to a node 
either as a switch function or a data center function but not both. For every 
$ u $ that is a request for a \ac{vnf} that can only be placed on a data center, 
$ \text{ms}_{u,v} $ is set to zero and $ \text{md}_{u,v} $ is set to one. 
This constraint is necessary for correctness of Constr.\,\ref{constr:m.md.p} and \ref{constr:m.ms.p} when the function has a non-zero value for both types of 
computational resource requirements (i.e., can be placed either on a switch node 
or on a data center node). Computational resource requirements of all requests 
mapped to a node should be less than or equal to available resources in that node.
In data center nodes only the data center resource requirements of the functions 
mapped to them are considered and in switch nodes only the switch resource 
requirements (Constr.\,\ref{constr:m.md.p} and \ref{constr:m.ms.p}).
Variables $ m $, $ \text{md} $, and $ \text{ms} $ are binary decision variables
and the product of two binary variables can easily be linearized. Therefore, 
Constr.\,\ref{constr:m.md.p}--\ref{constr:m.ms.p} can be considered as quadratic 
constraints instead of cubic. For every
function $ f $, up to $ n_{\text{inst}}(f) $ instances can be placed in the network
(Constr.\,\ref{constr:i<=inst}). Finally, Constr.\,\ref{constr:m<=req} ensures
that every instance of $ f $ handles no more than $ n_{\text{req}}(f) $ requests.

\subsubsection{Path Creation Constraints}
{\footnotesize
\begin{align}
    & \forall (v,v') {\in} E, \forall x,y {\in} V, \forall (u,u') {\in} U_\text{pairs} : \notag\\
    &\quad e_{v,v',x,y,u,u'} {\leq} m_{u,x} {\cdot} m_{u',y} \label{constr:e}
\end{align}
\begin{align}
    & \forall (u,u') {\in} U_\text{pairs} : \notag\\ 
    &\quad \sum_{(x,v) {\in} E, y {\in} V} e_{x,v,x,y,u,u'} {\cdot} m_{u,x} {\cdot} m_{u',y} {=} 1 \label{constr:start} \\
    &\quad \sum_{(x,v) {\in} E, y {\in} V} e_{x,v,x,y,u,u'} {\cdot} (1 {-} m_{u,x} {\cdot} m_{u',y}) {=} 0 \label{constr:notstart} \\
    &\quad \sum_{(v,y) {\in} E, x {\in} V} e_{v,y,x,y,u,u'} {\cdot} m_{u,x} {\cdot} m_{u',y} {=} 1 \label{constr:end} \\
    &\quad \sum_{(v,y) {\in} E, x {\in} V} e_{v,y,x,y,u,u'} {\cdot} (1 {-} m_{u,x} {\cdot} m_{u',y}) {=} 0 \label{constr:notend}\\
    & \forall (u,u') {\in} U_\text{pairs}, \forall w,x,y {\in} V : \notag\\
    &\quad \sum_{\substack{v {\in} V, v {\neq} y, \\ (v,w) {\in} E}} e_{v,w,x,y,u,u'} {=} \sum_{\substack{v' {\in} V, w {\neq} x,\\ (w,v') {\in} E}} e_{w,v',x,y,u,u'} \label{constr:preserve} \\  
    & \forall (u,u') {\in} U_\text{pairs}, \forall v,x,y {\in} V, x {\neq} y : \, e_{v,v,x,y,u,u'} {=} 0 \label{constr:loop1} \\
    & \forall (u,u') {\in} U_\text{pairs}, \forall x,y {\in} V, \forall (v,v'), (v',v) {\in} E, v {\neq} v' : \notag\\
    &\quad e_{v,v',x,y,u,u'} {+} e_{v',v,x,y,u,u'} {\leq} 1 \label{constr:loop2} \\
    & \forall (v,v') {\in} E : \notag\\
    &\quad \sum_{(u,u') {\in} U_\text{pairs},  \forall x,y {\in} V} e_{v,v',x,y,u,u'} {\cdot} d_\text{req}(u,u') {\leq} d(v,v') \label{constr:e.dreq} \\
    & \forall (a,a') {\in} A_\text{pairs} : \notag\\
    &\quad \sum_{\substack{(v,v') {\in} E, x,y {\in} V, \\ (u,u') {\in} \text{paths}(a,a')}} e_{v,v',x,y,u,u'} {\cdot} l(v,v') {\leq} l_\text{req}(a,a') \label{constr:e.lreq}   
\end{align}
}%
An edge in the network graph belongs to a path between nodes $ v $ and $ v' $ if 
there are requests mapped to these nodes and a path needs to be created 
between them (Constr.\,\ref{constr:e}). Constr.\,\ref{constr:start} ensures 
the path in network graph created for edge $ (u,u') $ in the \ac{vnf} graph starts 
at exactly one edge going out of node $ x $ in the network where request $ u $ is mapped
to. Without Constr.\,\ref{constr:notstart} any random edge might be marked as 
the first edge of this path. Analogously, Constr.\,\ref{constr:end}--\ref{constr:notend}
ensure the correctness and uniqueness of the creation of the last edge in the path.
In Constr.\,\ref{constr:start}--\ref{constr:notend}, the product of the binary
variables can be linearized to avoid having cubic constraints in the problem.
Constr.\,\ref{constr:preserve} ensures that for every node $ w $ in the network 
graph, if one of its incoming edges belongs to a path between the nodes where 
requests $ u $ and $ u' $ are mapped to, then one of its outgoing edges also 
belongs to this path. Excluded from this rule are the cases where the incoming 
edge to a node is the last edge in the path and where the outgoing edge from a 
node is the first edge in the path. Constraints \ref{constr:loop1}--\ref{constr:loop2}
prevent the creation of infinite loops and unnecessary extensions of the created paths.
For every edge in the network, the sum of the required data rates of all paths going
through that edge should be less than or equal to the data rate capacity of this edge (Constr.\,\ref{constr:e.dreq}).
Moreover, the sum of latencies of all edges that belong to a path between the start 
and end points of a flow should not exceed the maximum tolerable latency given for that flow
(Constr.\,\ref{constr:e.lreq}).

\subsubsection{Metrics Calculation Constraints}
{\footnotesize
\begin{align}
  & \forall v {\in} V : \sum_{f {\in} F} i_{f,v} {\leq} \mathcal{M'} {\cdot} \text{used}_{v}  \label{constr:use1} \\
  & \forall v {\in} V : \text{used}_{v} {\leq} \sum_{f {\in} F} i_{f,v}  \label{constr:use2} \\
  & \forall (v,v') {\in} E : \notag\\
  & \quad \text{remdr}_{v,v'} {=} d(v,v') {-} \sum_{\substack{(u,u') {\in} U_\text{pairs}, \\ \forall x,y {\in} V}} e_{v,v',x,y,u,u'} {\cdot} d_\text{req}(u,u') \label{constr:rdr} \\
  & \forall (u,u') {\in} U_\text{pairs} : \text{lat}_{u,u'} {=} \sum_{x,y {\in} V, (v,v') {\in} E} e_{v,v',x,y,u,u'} {\cdot} l(v,v')  \label{constr:lat}
\end{align}
}%
Using Constr.\,\ref{constr:use1} and \ref{constr:use2} we mark a network node
as \emph{used} if there is an instance of at least one function mapped to it.
$ \mathcal{M'} {\in} \mathbb{N} $ is a number larger than the sum on the left side 
of the inequality in Constr.\,\ref{constr:use1}. For each edge $ (v,v') $ in the
network, the remaining data rate after the placement of chained functions is calculated
by subtracting the sum of required data rates of all paths that go through this 
edge from the initial data rate of it (Constr.\,\ref{constr:rdr}). For every
edge $ (u,u') $ in the \ac{vnf} graph, the latency of the paths created between the 
nodes where requests $ u $ and $ u' $ are mapped to is equal to the sum of latencies
of all network edges that belong to this path (Constr.\,\ref{constr:lat}).

\subsection{Objectives}
\label{subsec:objective}

Different objectives can be targeted for placement optimization, and
each of them can result in a different mapping of the \ac{vnf} graphs into the 
network graph. We define three objective functions and describe 
the behavior of the placement process using each objective.

\subsubsection{Maximizing the remaining data rate on network links}
\vspace{-1mm}
{\footnotesize
\begin{equation}
 \text{maximize} \sum_{(v,v') {\in} E, v {\neq} v'} \text{remdr}_{v,v'}
\vspace{-1mm}
\end{equation}%
}%
As highly utilized links can result in congestion in the network, solutions that
leave more capacity on the links are desirable. This objective
aims at leaving more data rate on the links. By maximizing the sum of remaining
data rate over all edges except self-loops, it forces the placement 
algorithm to use self-loops (i.e., links between two functions that
are placed on one network node) more than other links.
 
\subsubsection{Minimizing the number of used nodes in the network}
\vspace{-1mm}
{\footnotesize
\begin{equation}
 \text{minimize} \sum_{v {\in} V } \text{used}_{v}
 \vspace{-1mm}
\end{equation}
}%
This objective can result in an energy-efficient solution by allowing more unused 
nodes to be switched off. However, it might concentrate the placement
of functions on a small subset of nodes causing congestion in the network.

\subsubsection{Minimizing the latency of the created paths}
\vspace{-1mm}
{\footnotesize
\begin{equation}
 \text{minimize} \sum_{(a,a') {\in}  l_\text{req}} \Bigl( \sum_{P {\in} \text{paths}(a,a')} \Bigl( \sum_{(u,u') {\in} P} \text{lat}_{u,u'} \Bigr) \Bigr) 
 \vspace{-1mm}
\end{equation}
}%
In complex chaining scenarios with branches in the structure, there are multiple simple
paths between the start and end points. As each path consists of different sets of
edges, they can have different latencies. This objective function minimizes the
mean latency of all paths created for all deployment requests.

These objective functions cause the placement to 
focus on a specific goal. For example, using the third objective, we get solutions
with minimum latency but the remaining data rate of the links or the number
of used nodes in the network are not predictable. There can be conflicts in the 
solutions that are considered as \emph{optimal} using each of these objectives. 
Results of our Pareto analysis (Section~\ref{sec:evaluation}) show that the three 
metrics can have trade-offs but are not necessarily conflicting.

\section{Evaluation}
\label{sec:evaluation}
We have performed two types of evaluation of our model and placement optimization process:
\begin{inparaenum}
 \item Observing the behavior of the placement process and our heuristic for 
 reducing the runtime of the process when there are several ordering 
 possibilities in deployment requests;
 \item Pareto analysis for showing the possible trade-offs between our metrics of
 interest.
\end{inparaenum}
Currently, there is no commonly accepted evaluation model
for either the actual chained network functions or the user requests for tenant
applications with chains of virtual or physical network functions in their structure.
Therefore, we have designed small evaluation scenarios with 
manually created deployment requests on a substrate network with 12 nodes
and 42 directed edges (including self-loops), based on the \emph{abilene} network
from SNDlib \cite{SNDlib10}. We have built these requests to test the capabilities
of the request processing and placement steps while being compatible with known
use cases of chaining \acp{vnf}~\cite{draft-liu-sfc-use-cases-05, leymannslides}. 
We have used the Gurobi Optimizer to solve the 
\ac{miqcp} on machines with Intel Xeon X5650 CPUs running at 2.67\,GHz. 

\subsection{Evaluation of Optional Orders in Chaining Requests}

For this part, we have used a set of chaining requests which allow an arbitrary order
among a set of functions. Our placement results show that for small requests that
have small requirements compared to the available resources in the network sorting
the functions according to their ratios of outgoing to incoming data rate gives 
the best (or one of the best) solutions for the requests. When placing several requests 
the combination of requests have higher requirements and there can be dependencies 
between different requests (e.g., shared \acp{vnf}). In that case, combining the 
\emph{sorted} chains may result in a sub-optimal solution. 
The runtime, however, should also be considered. For example, 
in our evaluation settings, placing a combination of 3 requests with optional orders 
(resulting in 6, 6, and 4 \ac{vnf} graphs, respectively) 
by sorting them takes an average of 13 minutes using different objective 
functions to give a close to optimal solution. But 
computing all combinations ($ 6 {\cdot} 6 {\cdot} 4 {=} 144 $) to find the optimal one needs 31 hours. For 
this example, placement of the sorted chains was one of the optimal placements regarding 
remaining data rate and the latency but it used more network nodes compared to some
other solutions. Our heuristic offers the opportunity to place the chained functions in an
acceptable time by allowing a slight deviation from the optimal solution. Considering
the fact that even a simple set with 3 requests can result in 144 combinations, 
we use this heuristic for the second part of our evaluations and choose the
sorted chain for each of the requests in our request sets.

\subsection{Pareto Set Analysis}

\begin{figure}[!t]
\centerline{\subfloat[Results for a sample request set with optimal solution]{\includegraphics[width=0.77\linewidth]{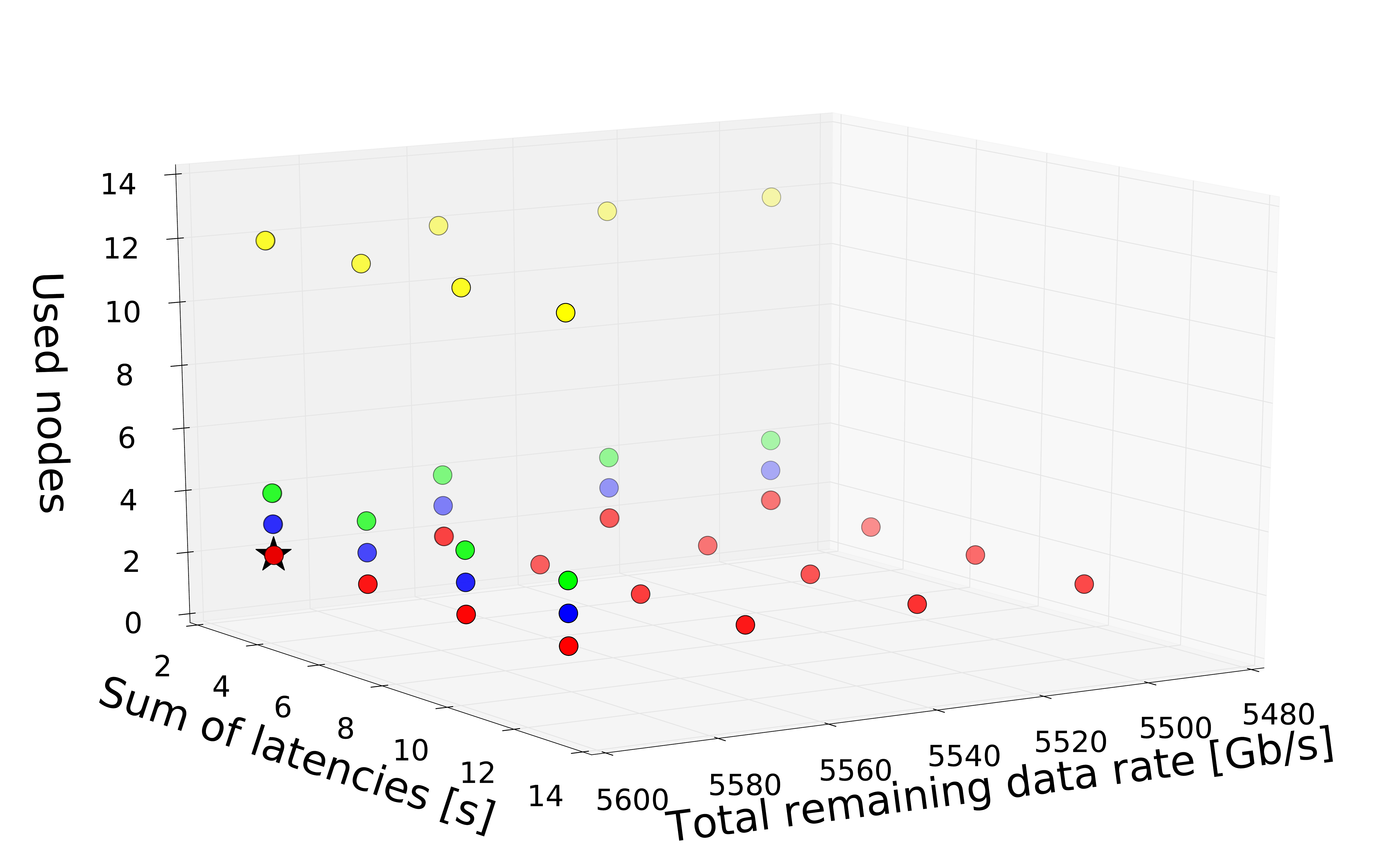}
\label{fig:xsdatacenter}}}
\vspace{-12pt}
\centerline{\subfloat[Results for a sample request set including a Pareto set]{\includegraphics[width=0.77\linewidth]{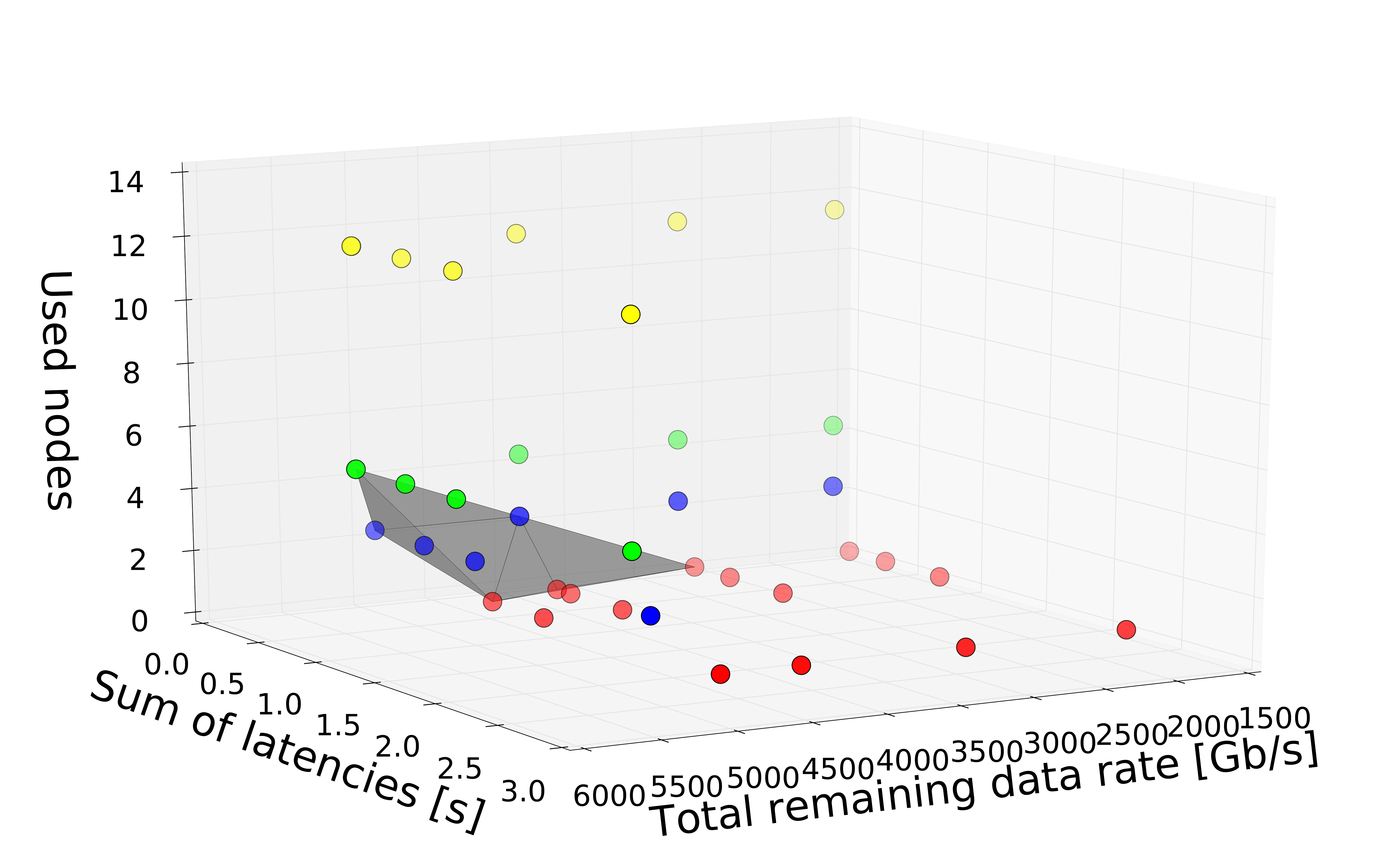}
\label{fig:xsbroadband}}}
\caption{Sample results from the multi-objective Pareto analysis}
\label{fig:pareto}
\vspace{-2mm}
\end{figure}

For the Pareto analysis, we have performed 
the placement for different sets of deployment requests using the objectives defined in
Section~\ref{subsec:objective}. Our request sets resemble three different \ac{vnf} chaining
scenarios (broadband, mobile core, and data center networks) from the
IETF draft on service function chaining use cases \cite{draft-liu-sfc-use-cases-05}. We 
have defined two different sets of requests with different complexities for each 
scenario. First we did a range estimation run for the request sets by performing the 
placement using each of the objectives and recording the highest and lowest values 
for the metrics. After identifying the interesting ranges, we performed
the Pareto optimization. Some of our results show trade-offs between optimizing
the remaining data rate, number of used nodes, and latency; there are also
results that show it is possible to find a placement that optimizes all three
metrics. Figure~\ref{fig:pareto} shows the results for two of our request sets.
For better visibility, we use different colors for different number of used nodes 
in this figure. As illustrated in Figure~\ref{fig:xsdatacenter}, the objectives are not 
always conflicting. If there are enough free resources in the network, it might be possible 
to find a solution that is optimal in terms of all three metrics (the solution marked by a star).

Results of the Pareto optimization for another one of our request sets is shown 
in Figure~\ref{fig:xsbroadband}. Allowing the placement to use more
nodes gives results with lower latency and higher remaining data rate. But 
after a certain point increasing the number of used nodes does not improve the 
results anymore. The figure also 
shows a trade-off between latency and remaining data rate. Using a fixed number of 
nodes, the latency of the solution increases to get a higher value for the remaining 
data rate, which also means that improving the latency can result in 
a lower remaining data rate.

Network operators can have different objectives which require 
different placement solutions. Results of our Pareto analysis can help the operators 
prioritize their optimization goals and choose the right objective functions. 

\section{Conclusion}
\label{sec:conclusion}

In this paper, we have presented a formal model for specifying \ac{vnf} 
chaining requests, including a context-free language for denoting complex 
composition of \ac{vnf}. Using this model, we have formulated an optimization 
problem for placing the chained \ac{vnf} in an operator's network with multiple 
sites, based on requirements of the tenants and the operator. Our evaluations have shown that 
placement of chained \ac{vnf} is not a trivial problem and that placement decisions 
have to be taken differently according to the desired placement objective (i.e. 
remaining data rate, latency, number of used network nodes). Our results warrant 
further investigation to allow fast and efficient placement of \acp{vnf} and to 
facilitate the deployment of \ac{nfv} in different types of networks.

\begin{acronym}[WWWW]

\acro{adt}[ADT]{Application Deployment Toolkit}
\acro{alg}[ALG]{Application-Level Gateway}
\acro{av}[AV]{Anti-virus}

\acro{bnf}[BNF]{Backus-Naur Form}
\acro{bng}[BNG]{Broadband Network Gateway}


\acro{cache}[CACHE]{Cache}
\acro{cr}[CR]{Core Router}

\acro{dpi}[DPI]{Deep Packet Inspector}

\acro{etsi}[ETSI]{European Telecommunications Standards Institute}

\acro{fpga}[FPGA]{Field Programmable Gate Array}
\acro{fw}[FW]{Firewall}

\acro{ggsn}[GGSN]{Gateway GPRS Support Node}

\acro{http}[HTTP]{Hypertext Transfer Protocol }

\acro{ids}[IDS]{Intrusion Detection System}
\acro{ietf}[IETF]{Internet Engineering Task Force}
\acro{ip}[IP]{Internet Protocol}
\acro{isg}[ISG]{Industry Specification Group}



\acro{lb}[LB]{Load Balancer}


\acro{miqcp}[MIQCP]{Mixed Integer Quadratically Constrained Program}

\acro{nat}[NAT]{Network Address Translator}
\acro{nf}[NF]{Network Function}
\acro{nfv}[NFV]{Network Function Virtualization}


\acro{pctl}[PCTL]{Parental Control}
\acro{pgw}[PGW]{Packet Data Network Gateway}


\acro{reg}[REG]{Regional Network}

\acro{sdn}[SDN]{Software-Defined Networking}
\acro{srv}[SRV]{Server}



\acro{vm}[VM]{Virtual Machine}
\acro{vnf}[VNF]{Virtual Network Function}
\acro{vopt}[VOPT]{Video Optimizer}

\acro{wap}[WAP]{Wireless Application Protocol}
\acro{wapgw}[WAPGW]{Wireless Application Protocol Gateway}
\acro{wopt}[WOPT]{Wide-Area Network Optimizer}




\end{acronym}

\bibliographystyle{IEEEtran}
\bibliography{library,extra}

\begin{thebibliography}{10}
\providecommand{\url}[1]{#1}
\csname url@samestyle\endcsname
\providecommand{\newblock}{\relax}
\providecommand{\bibinfo}[2]{#2}
\providecommand{\BIBentrySTDinterwordspacing}{\spaceskip=0pt\relax}
\providecommand{\BIBentryALTinterwordstretchfactor}{4}
\providecommand{\BIBentryALTinterwordspacing}{\spaceskip=\fontdimen2\font plus
\BIBentryALTinterwordstretchfactor\fontdimen3\font minus
  \fontdimen4\font\relax}
\providecommand{\BIBforeignlanguage}[2]{{%
\expandafter\ifx\csname l@#1\endcsname\relax
\typeout{** WARNING: IEEEtran.bst: No hyphenation pattern has been}%
\typeout{** loaded for the language `#1'. Using the pattern for}%
\typeout{** the default language instead.}%
\else
\language=\csname l@#1\endcsname
\fi
#2}}
\providecommand{\BIBdecl}{\relax}
\BIBdecl

\bibitem{Carpenter2002}
B.~Carpenter and S.~Brim, ``{Middleboxes: Taxonomy and Issues. RFC 3234},''
  2002.

\bibitem{schneiderstandardizations}
F.~Schneider, T.~Egawa, S.~Schaller, S.-i. Hayano, M.~Sch{\"o}ller, and
  F.~Zdarsky, ``{Standardizations of SDN and Its Practical Implementation}.''

\bibitem{draft-ietf-sfc-problem-statement-05}
P.~Quinn and T.~Nadeau, ``{Service Function Chaining Problem Statement},''
  {Active Internet-Draft}, {IETF Secretariat}, {Internet-Draft}
  draft-ietf-sfc-problem-statement-05, Apr. 2014.

\bibitem{Sekar2011}
V.~Sekar, S.~Ratnasamy, M.~K. Reiter, N.~Egi, and G.~Shi, ``{The middlebox
  manifesto: enabling innovation in middlebox deployment},'' \emph{Proceedings
  of the 10th ACM Workshop on Hot Topics in Networks}, p.~21, 2011.

\bibitem{Introduction2012}
R.~Guerzoni, ``{Network Functions Virtualisation: An Introduction, Benefits,
  Enablers, Challenges and Call for Action, Introductory white paper},''
  \emph{SDN and OpenFlow World Congress}, 2012.

\bibitem{draft-xjz-nfv-model-problem-statement-00}
W.~Xu, Y.~Jiang, and C.~Zhou, ``{Problem Statement of Network Functions
  Virtualization Model},'' {Internet-Draft}, {IETF Secretariat},
  {Internet-Draft} draft-xjz-nfv-model-problem-statement-00, Sep. 2013.

\bibitem{draft-liu-sfc-use-cases-05}
W.~Liu, H.~Li, O.~Huang, M.~Boucadair, N.~Leymann, Z.~Cao, Q.~Sun, and C.~Pham,
  ``{Service Function Chaining (SFC) Use Cases},'' {Active Internet-Draft},
  {IETF Secretariat}, {Internet-Draft} draft-liu-sfc-use-cases-05, Apr. 2014.

\bibitem{draft-boucadair-sfc-framework-02}
M.~Boucadair, C.~Jacquenet, R.~Parker, D.~Lopez, J.~Guichard, and C.~Pignataro,
  ``{Service Function Chaining: Framework and Architecture},'' {Active
  Internet-Draft}, {IETF Secretariat}, {Internet-Draft}
  draft-boucadair-sfc-framework-02, Feb. 2014.

\bibitem{Nagy2007}
G.~Nagy and S.~Salhi, ``{Location-routing: Issues, models and methods},''
  \emph{European Journal of Operational Research}, vol. 177, no.~2, pp.
  649--672, Mar. 2007.

\bibitem{Prodhon2014}
C.~Prodhon and C.~Prins, ``{A Survey of Recent Research on Location-Routing
  Problems},'' \emph{European Journal of Operational Research}, Jan. 2014.

\bibitem{Fischer2013}
A.~Fischer, J.~F. Botero, M.~T. Beck, H.~de~Meer, and X.~Hesselbach, ``{Virtual
  Network Embedding: A Survey},'' \emph{IEEE Communications Surveys \&
  Tutorials}, vol.~15, no.~4, pp. 1888--1906, 2013.

\bibitem{Fuerst2013}
C.~Fuerst, S.~Schmid, and A.~Feldmann, ``{Benefits and Limitations of
  Pre-Clustering},'' in \emph{IEEE 2nd International Conference on Cloud
  Networking (CloudNet 2013)}, San Francisco, CA, 2013, pp. 91--98.

\bibitem{Joseph2008}
D.~Joseph and I.~Stoica, ``Modeling middleboxes,'' \emph{Network, IEEE},
  vol.~22, no.~5, pp. 20--25, 2008.

\bibitem{gember2013stratos}
A.~Gember, A.~Krishnamurthy, S.~S. John, R.~Grandl, X.~Gao, A.~Anand,
  T.~Benson, A.~Akella, and V.~Sekar, ``Stratos: A network-aware orchestration
  layer for middleboxes in the cloud,'' \emph{arXiv preprint arXiv:1305.0209},
  2013.

\bibitem{SNDlib10}
S.~Orlowski, M.~Pi{\'o}ro, A.~Tomaszewski, and R.~Wess{\"a}ly,
  ``\BIBforeignlanguage{English}{{SNDlib} 1.0--{S}urvivable {N}etwork {D}esign
  {L}ibrary},'' in \emph{\BIBforeignlanguage{English}{Proceedings of the 3rd
  International Network Optimization Conference (INOC 2007), Spa, Belgium}},
  April 2007.

\bibitem{leymannslides}
N.~Leymann, ``{Flexible Service Chaining. Requirements and Architectures.}''
  \url{http://ewsdn.org/presentations/Presentations_2013/EWSDN-2013-v10a.pdf},
  {2013}.

\end{thebibliography}
%
%
%

\end{document}